\documentclass[times,twocolumn,final]{elsarticle}
\usepackage{prletters}
\usepackage{multirow}
\usepackage{comment}
\usepackage[colorlinks]{hyperref}
\usepackage{subfigure}
\usepackage{amssymb}
\usepackage{amsmath}
\usepackage{natbib}
\usepackage{float}
\usepackage{booktabs}

\journal{Pattern Recognition Letters}

\begin{document}

\begin{frontmatter}

\title{Zero-Shot KWS for Children's Speech using Layer-Wise Features from SSL Models}

\author[*]{Subham Kutum}
\ead{phec240001@nitsikkim.ac.in}

\author[*]{Abhijit Sinha}
\ead{phec230023@nitsikkim.ac.in}

 \author[*]{Hemant Kumar Kathania}
\ead{hemant.ece@nitsikkim.ac.in}

\author[**]{Sudarsana Reddy Kadiri}
\ead{skadiri@usc.edu}

\author[*]{Mahesh Chandra Govil}
\ead{director@nitsikkim.ac.in}

\affiliation[*]{organization={National Institute of Technology, Sikkim},
            addressline={Ravangla}, 
            postcode={737139}, 
            state={Sikkim},
            country={India}}

\affiliation[**]{Signal Analysis and Interpretation Laboratory, University of Southern California, Los Angeles, USA}

\begin{abstract}
Numerous methods have been proposed to enhance Keyword Spotting (KWS) in adult speech, but children's speech presents unique challenges for KWS systems due to its distinct acoustic and linguistic characteristics. This paper introduces a zero-shot KWS approach that leverages state-of-the-art self-supervised learning (SSL) models, including Wav2Vec2, HuBERT and Data2Vec. Features are extracted layer-wise from these SSL models and used to train a Kaldi-based DNN KWS system. The WSJCAM0 adult speech dataset was used for training, while the PFSTAR children's speech dataset was used for testing, demonstrating the zero-shot capability of our method. Our approach achieved state-of-the-art results across all keyword sets for children's speech. Notably, the Wav2Vec2 model, particularly layer 22, performed the best, delivering an ATWV score of 0.691, a MTWV score of 0.7003 and probability of false alarm ($P_{\text{fa}}$) and probability of miss ($P_{\text{miss}}$) of 0.0164 and 0.0547 respectively, for a set of 30 keywords. Furthermore, age-specific performance evaluation confirmed the system’s effectiveness across different age groups of children. To assess the system's robustness against noise, additional experiments were conducted using the best-performing layer of the best-performing Wav2Vec2 model. The results demonstrated a significant improvement over traditional MFCC-based baseline, emphasizing the potential of SSL embeddings even in noisy conditions. To further gerneralize the KWS framework, the experiments were repeated for an additional CMU dataset. Statistical analyses including paired t-tests and Wilcoxon signed-rank tests were also performed, confirming that the observed improvements are statistically significant and further validating the reliability of the proposed framework. Overall the results highlight the significant contribution of SSL features in enhancing Zero-Shot KWS performance for children's speech, effectively addressing the challenges associated with the distinct characteristics of child speakers.

\end{abstract}


\end{frontmatter}

\section{Introduction}
\label{sec:Introduction}
Keyword Spotting (KWS) involves detecting specific words or phrases within spoken language. It plays a crucial role in various applications, such as voice-activated devices, smart home systems, security systems and automated transcription services. As the demand for voice-controlled applications grows, the significance of KWS has increased substantially. These systems reduce the need for manual input, allowing users to quickly and accurately access information with minimal effort.

Traditionally, KWS systems have been trained on adult speech data using methods that range from basic keyword models applying likelihood ratios for keyword identification \cite{weintraub1995lvcsr, szoke2005comparison}, to more advanced techniques such as phoneme-based recognition \cite{rose1990hidden} and Large Vocabulary Continuous Speech Recognition (LVCSR) systems with language models that search for keywords within an ASR lattice \cite{can2011lattice}. Recent developments in deep learning have greatly enhanced the performance and dependability of KWS systems \cite{hwang2015online,lopez2021deep}. These advancements have enabled KWS systems to operate effectively in more challenging audio scenarios while improving keyword detection accuracy. Mazumder et al. \cite{mazumder2021few} presented a method leveraging few-shot transfer learning to address KWS tasks across various languages. Similarly, \cite{berg2021keyword} developed a transformer-based framework for KWS, demonstrating state-of-the-art performance on the Google Speech Commands \cite{warden2018speech} dataset. Moreover, various KWS systems based on self-supervised learning (SSL) have been developed. Hussain et al. \cite{hussain2022multi}, proposed a fine-tuned single end-to-end Wav2Vec2 model using multi-task learning. Their framework jointly optimizes for both KWS and speaker identification, and requires labeled task-specific data for supervised training. Mork et al. \cite{mork2024noise}, proposed a student–teacher architecture based on Data2Vec. In their method, the model is trained on clean, noisy or denoised MFCC features, and a downstream multilayer perceptron (MLP) is trained for keyword classification. Their approach involves significant supervised fine-tuning and domain adaptation. SSL methods generate pseudo-targets from the data itself, enabling the model to learn robust representations of the data domain without the need for labeled annotations \cite{baevski2020wav2vec, chen2022wavlm}. These models have proven effective in learning strong representations from unlabeled data, yielding impressive performance across various downstream tasks, including KWS. 
A recent study by Zhu et al. \cite{zhu2024ge2e} proposed a fully end-to-end, zero‐shot GE2E-KWS. Huang et al. \cite{huang2023building} proposed KWS framework which is built on top of end-to-end (E2E) ASR systems using N-best lists.

Presently, KWS systems designed for adult speech have made significant progress. However, the development of KWS systems specifically tailored for children has received relatively little attention. The unique characteristics of children's speech, such as higher pitch and distinct pronunciation patterns pose significant challenges for accurate keyword recognition. Moreover deploying KWS systems for children introduces unique privacy concerns. KWS for children need large amounts of labeled child speech recordings, which are inherently sensitive. Collecting and storing this data creates risks of unauthorized access, misuse of recordings. One way to mitigate this challenge is a zero-shot KWS models trained on adult speech that can avoid these risks by eliminating the need to collect and store children’s voice data.

Recently, SSL models have made significant strides in various speech processing tasks, enabling the models to learn robust speech representations from large amounts of unlabeled audio \cite{radford2023robust}. Studies have demonstrated the effectiveness of these models for various speech tasks involving children \cite{sinha2024effect,Li2024AnalysisOS}. Zero-shot KWS is crucial for advancing speech technology, particularly in scenarios where labeled data is scarce or unavailable. The significance of zero-shot KWS lies in its ability to detect previously unseen keywords without requiring task-specific training. A few studies have explored KWS in low-resource scenarios, including \cite{yu2023few} and \cite{jacobs2024multilingual}.
Moreover the real-world deployment of KWS systems is often hindered by the challenges posed by background noises. In environments such as public spaces, homes, or vehicles, ambient noise can obscure keyword features, leading to false activations or missed detections. For example, in automotive settings, engine noise, road sounds, and passenger conversations can degrade system performance, reducing reliability in applications like voice-controlled navigation or call management. In \cite{mork2024noise}, authors introduced a Data2vec model to enhance the noise robustness of KWS.

This study introduces a novel zero-shot KWS framework for children's speech using features extracted from SSL models. We explore features extracted from 25 layers of three state-of-the-art pretrained SSL models: Wav2Vec2, HuBERT and Data2Vec \cite{baevski2020wav2vec, 9585401,baevski2022data2vec} for a Zero-Shot KWS designed for children's speech, a study which was not explored before. The proposed KWS system is trained on the WSJCAM0 adult speech dataset and evaluated on the PFSTAR children's speech dataset, demonstrating its Zero-Shot capabilities. This study determines the optimal layer features for effective children KWS in a zero-shot setup and further conducts an age-specific analysis across different groups to evaluate the influence of speech characteristics on system performance. To assess the system's robustness to noise, comprehensive experiments were performed under various noise conditions at SNR levels of 5, 10 and 15 dB. The noise types included Babble, Volvo, Factory and White noise from the NOISEX-92 dataset, along with ambulance siren, crowd, thunderstorm and birds chirping noises from the MUSAN noise dataset. To further generalize the proposed framework, additional experiments were conducted on the CMU Kids dataset using a similar zero-shot setup.

The rest of the paper is structured as follows: Section \ref{sec:sec2} outlines the proposed framework, detailing the architecture and methodologies employed. Section \ref{sec:Database and Experimental} provides a comprehensive details of the database and experimental setup. Section \ref{sec:sec4} presents the results and discussion. Finally, Section  \ref{sec:sec5} concludes with a summary of findings and suggestions for future work.

\section{Proposed frame work} 
\label{sec:sec2}
\begin{figure}[!h]
 \centering
        \includegraphics[width=7cm]{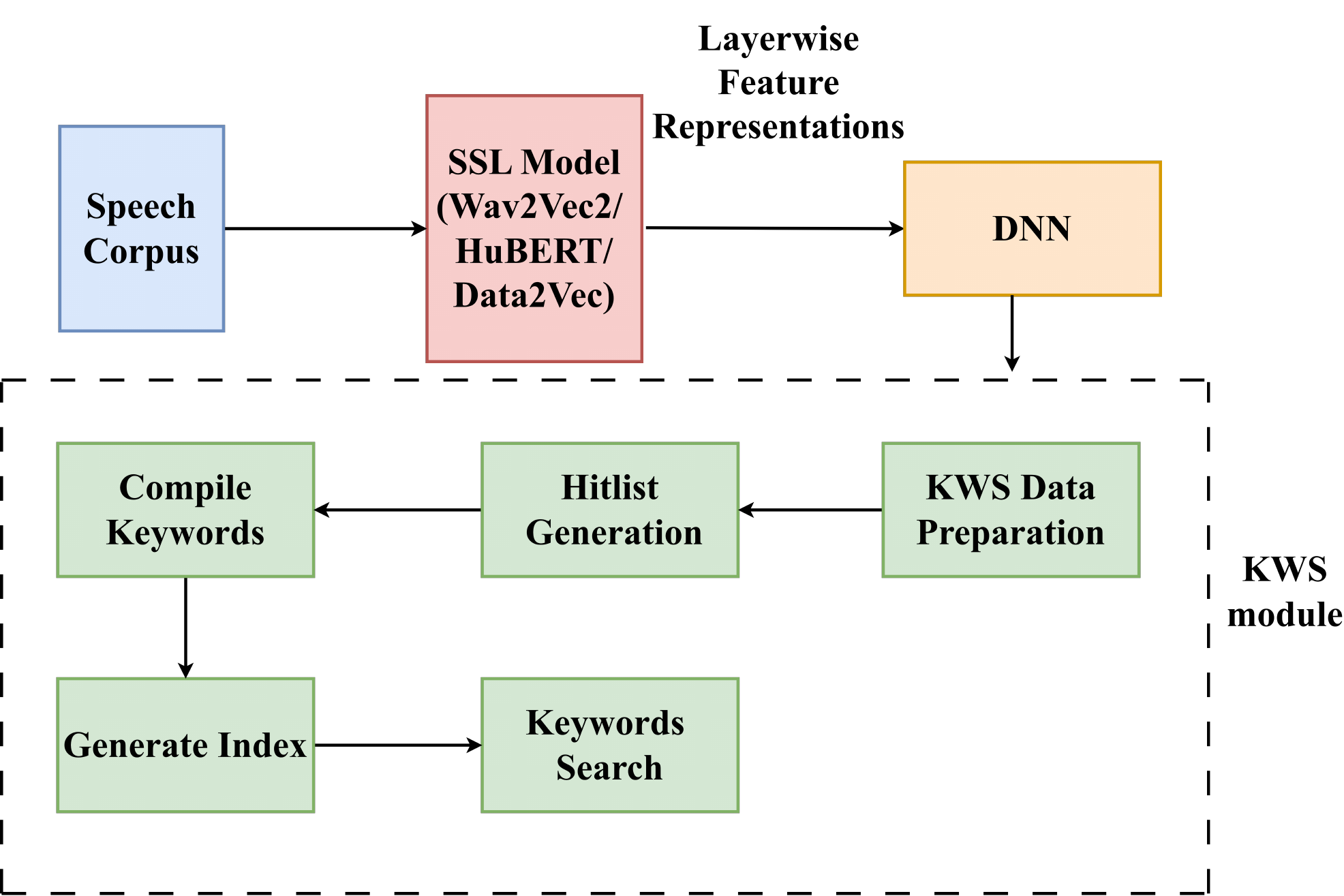} 
        \vspace{-0.2cm}
\caption{ Block diagram illustrating the proposed keyword spotting (KWS) framework for children’s speech. Features are extracted from all the layers of the SSL model to build a DNN system and then KWS module is applied upon it.}
         \label{fig:framework}
         \vspace{-0.4cm}
\end{figure}

The proposed framework, illustrated in Figure \ref{fig:framework}, outlines the architecture of a Zero-Shot KWS system for children's speech, leveraging layer-wise features from self-supervised learning (SSL) models. The system incorporates feature extraction from state-of-the-art SSL models: Wav2Vec2, HuBERT and Data2Vec followed by a Kaldi-based KWS pipeline. Each SSL model outputs 1024-dimensional features across 25 hidden layers: the initial layer (indexed as 0) captures representations from a convolutional neural network (CNN) feature encoder, while the remaining 24 layers (indexed 1 to 24) are transformer encoders. These models differ in their training objectives and the nature of their learned representations.

Wav2Vec2 \cite{baevski2020wav2vec} utilizes a contrastive loss to produce robust, noise-invariant embeddings in the upper layers. HuBERT \cite{9585401} employs iterative masked prediction over clustered units, resulting in phoneme-like representations in the mid-layers and progressively abstract features in deeper layers. Data2Vec \cite{baevski2022data2vec} adopts a teacher-student paradigm, guiding the model to predict contextualized latent targets, thereby learning smooth, semantically enriched representations. By extracting features from all 25 layers, the framework captures both fine-grained phonetic detail and higher-level contextual semantics, which are then input to a deep neural network (DNN) acoustic model for temporal alignment, lattice generation, and keyword detection.

The downstream KWS process begins with data preparation, including utterance-to-speaker mappings, trial list construction, and keyword indexing. During hitlist generation, test utterances are decoded using the DNN model to identify potential keyword candidates. These candidates are then compiled into Finite State Transducers (FSTs), representing alternative pronunciations of each keyword. An index is constructed over the FSTs to enable efficient search without redundant processing. Finally, the search and scoring module conducts FST-based keyword matching across the lattices, normalizes detection scores, and evaluates performance under zero-shot conditions. This architecture demonstrates the effectiveness of leveraging multi-layer SSL representations for keyword detection in children's speech.

\section{Database and Experimental Setup}
\label{sec:Database and Experimental}

This section outlines the dataset employed and details the experimental configuration. To develop the KWS system, we adopted the standard Kaldi framework \cite{povey2011kaldi} {(https://github.com/kaldi-asr/kaldi)}, which leverages a lattice re-scoring technique for efficient keyword indexing and retrieval.

\vspace{-0.4cm}

\begin{table}[!htbp]
    \centering
    \footnotesize
\caption{List of keywords from the PFSTAR dataset used in the study.}
    \label{tab:keywords}
    \vspace{0.2cm}
   \begin{tabular}{p{2cm} p{5.5cm}} \hline 
         No. of Keywords & List of Keywords\\ \hline
         10 & ZERO, ONE, TWO, THREE, FOUR, FIVE, SIX, SEVEN, EIGHT, NINE \\ 
        
         20 & ZERO, ONE, TWO, THREE, FOUR, FIVE, SIX, SEVEN, EIGHT, NINE, {\color{blue} TEN, THERE, THEY, BANK, NUMBER, POINT, MONTH, WITH, YEAR, PEOPLE} \\ 
    
         30 & ZERO, ONE, TWO, THREE, FOUR, FIVE, SIX, SEVEN, EIGHT, NINE, {\color{blue} TEN, THERE, THEY, BANK, NUMBER, POINT, MONTH, WITH, YEAR, PEOPLE,}  {\color{magenta} GOT, ORANGE, BEAUTIFUL, LIKE, YELLOW, TEACHER, TEETH, BIRTHDAY, RED, FEBRUARY} \\ \hline
    \end{tabular}
\vspace{-0.4cm}    
\end{table}
\vspace{-0.3cm}
\subsection{Database}
\label{sec:Database}
In this study, we utilized two datasets: WSJCAM0 \cite{robinson1995wsjcamo} for training and PFSTAR \cite{batliner2005pf_star} for testing. The WSJCAM0 dataset consists of British English speech recordings from 140 adult speakers, with each speaker providing around 110 utterances. For training purposes, 15.5 hours of audio data from 92 speakers were utilized. The PFSTAR dataset, on the other hand, features speech recordings from children aged 4 to 14 years.  A subset of this dataset, containing 1.1 hours of audio from 60 speakers, including 28 females, was used for testing.

The proposed approach illustrates a zero-resource method for child keyword spotting, where adult speech data from WSJCAM0 is used for training and child speech data from PFSTAR for testing. Three sets of 10, 20, and 30 keywords, as listed in Table~\ref{tab:keywords}, were considered. The setup enables evaluation of KWS performance across different keyword sets and facilitates comparison with prior studies.

\begin{figure*}[!t]
    \centering
    \begin{minipage}[t]{0.39\textwidth}
        \centering
        \includegraphics[width=\textwidth,trim=0.1cm 0cm 0.40cm 0cm,clip]{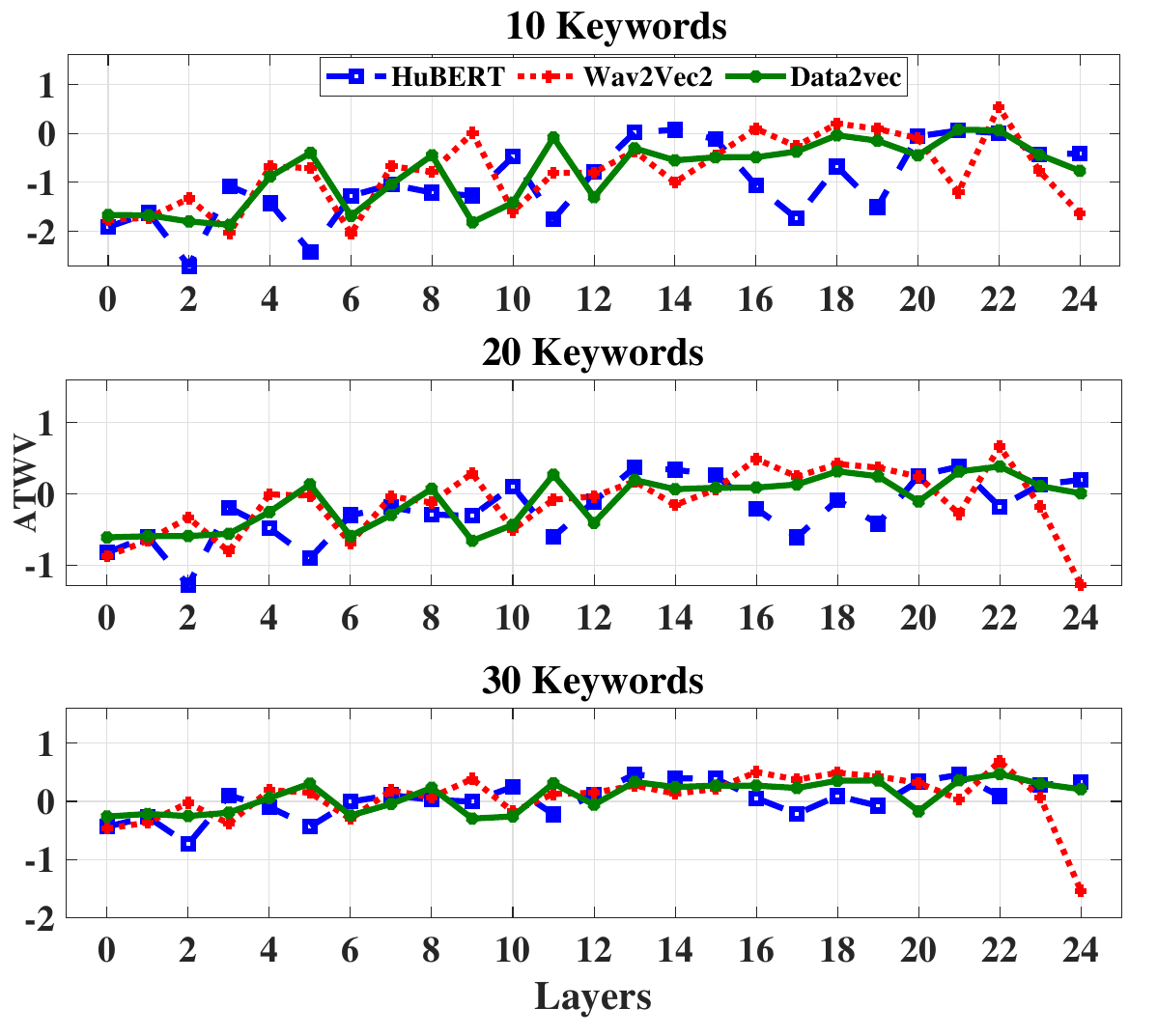}
    \end{minipage}
    \hspace{0.5cm}
    \begin{minipage}[t]{0.39\textwidth}
        \centering
        \includegraphics[width=\textwidth,trim=0.1cm 0cm 0.5cm 0cm,clip]{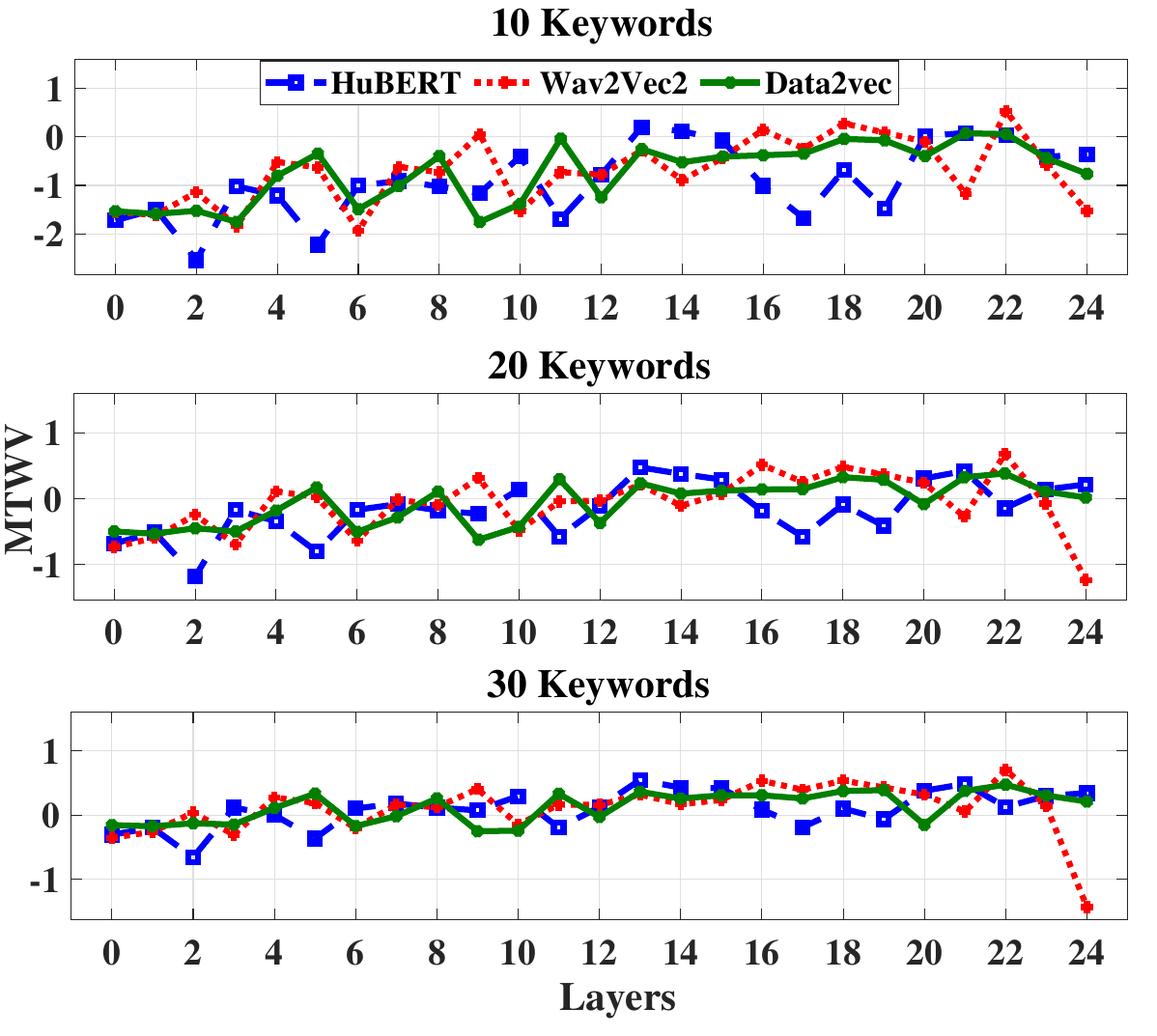}
    \end{minipage}
    \caption{ATWV and MTWV results for 10, 20 and 30 keywords using layer-wise features from SSL models: Wav2Vec2, HuBERT and Data2Vec, on the PFSTAR dataset.}
    \label{fig:atwv_mtwv}
    \vspace{3pt}
\end{figure*}

\begin{figure*}[!h]
    \centering
    \begin{minipage}[t]{0.39\textwidth}
        \centering
        \includegraphics[width=\textwidth,trim=0.1cm 0cm 0.40cm 0cm,clip]{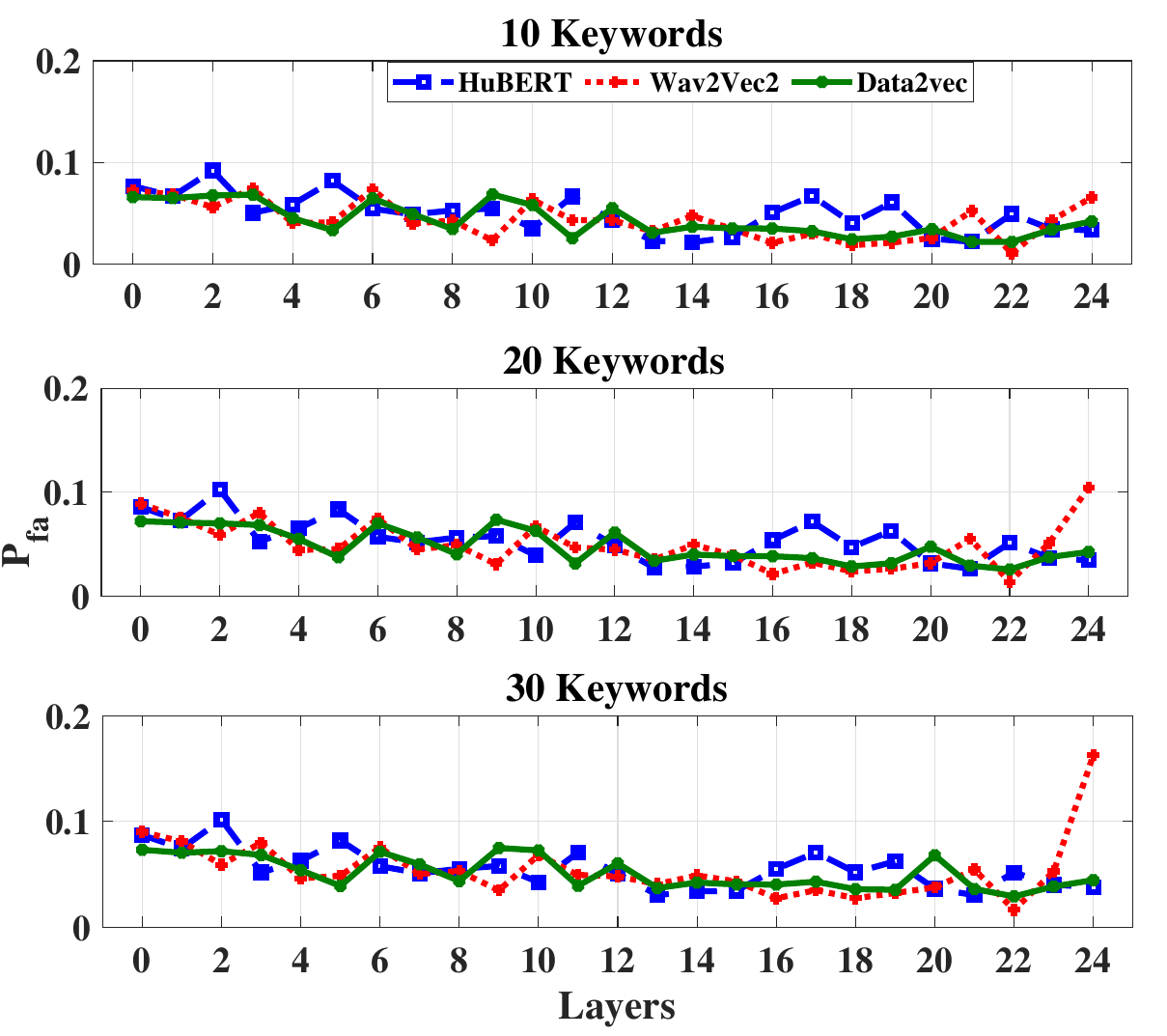}
    \end{minipage}
    \hspace{0.5cm}
    \begin{minipage}[t]{0.39\textwidth}
        \centering
        \includegraphics[width=\textwidth]{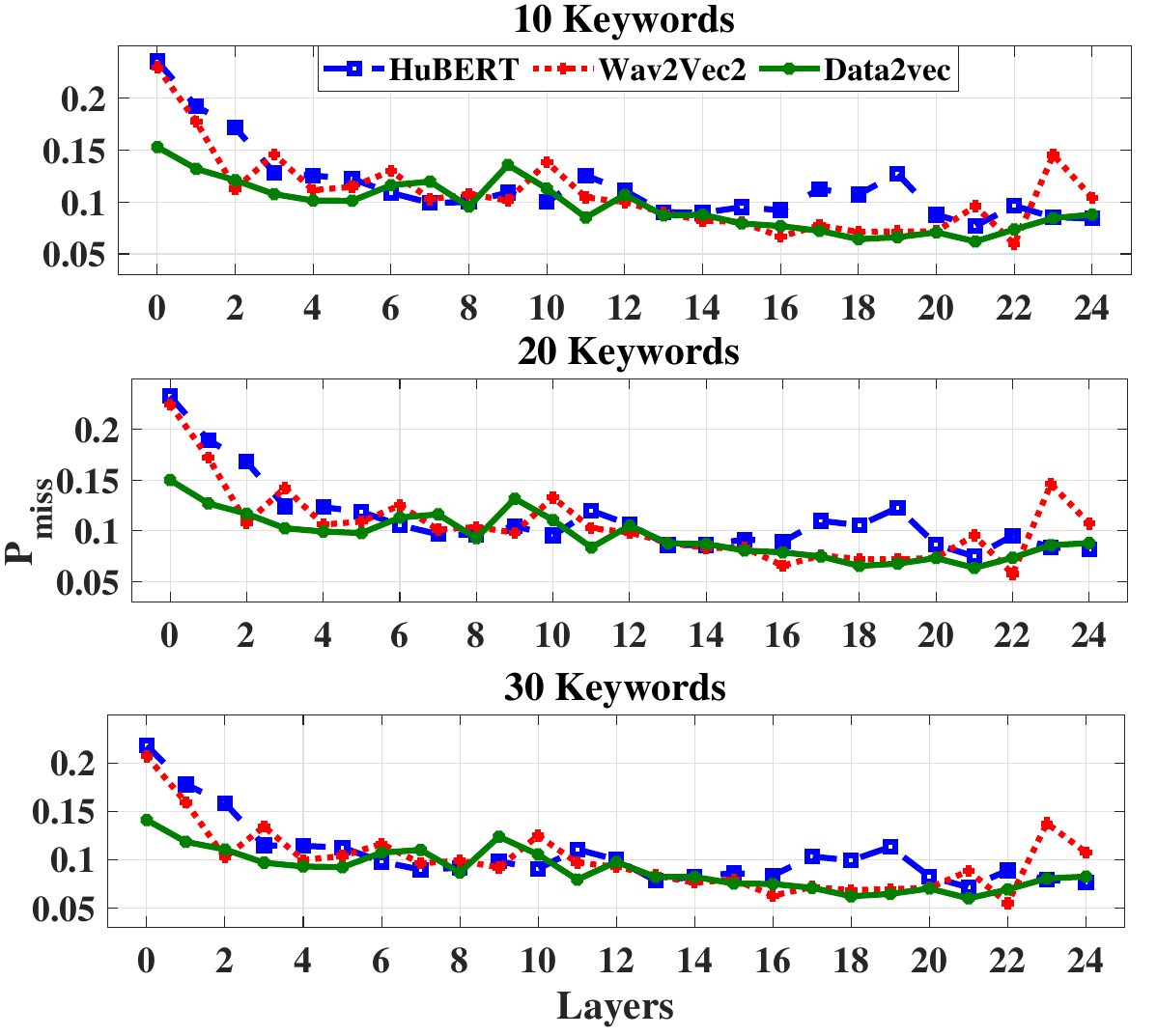}
    \end{minipage}
    \caption{$P_{\text{fa}}$ and $P_{\text{miss}}$ results for 10, 20 and 30 keywords using layer-wise features from SSL models: Wav2Vec2, HuBERT and Data2Vec, on the PFSTAR dataset.}
    \label{fig:pfa_pmiss}
    \vspace{-0.3cm}
\end{figure*}

\subsection{Kaldi based baseline KWS system}

The KWS system was developed using the KALDI toolkit. To extract MFCC feature vectors, the speech data underwent analysis with overlapping 20-ms Hamming-windowed frames and a frame shift of 10 ms. A 40-channel Mel filter bank was employed to derive 13-dimensional base MFCC features. These features were temporally spliced by appending four frames from both the preceding and succeeding contexts, resulting in a 117-dimensional feature vector. Dimensionality reduction to 40 was performed using linear discriminant analysis, followed by maximum likelihood linear transformation. Cepstral mean and variance normalization (CMVN) were applied for feature decorrelation and normalization was achieved using feature-space maximum likelihood linear regression (fMLLR). Speaker adaptive training was used to generate the fMLLR transformations for both training and test datasets. The acoustic models were trained using context-dependent hidden Markov models (HMMs). Gaussian mixture models (GMMs) and deep neural networks (DNNs) were utilized to compute the observation probabilities for the HMM states. A DNN-based acoustic model was subsequently employed to construct the KWS system.

\subsection{Self-Supervised Learning (SSL) Features}
In this study we employed three state-of-the-art SSL pre-trained models: Wav2Vec2, HuBERT and Data2Vec to extract layer-wise features. Although these models employ different training approaches, they share a common architectural framework. Each model uses convolutional neural networks (CNNs) to process raw audio into latent representations, capturing local acoustic features from the waveform. These CNN-extracted features are then fed into Transformer encoders, which captures long-range dependencies. Each model feature a total of 25 hidden layers. The first hidden layer represents the CNN output, while the subsequent 24 layers are Transformer layers that enhance contextual information.

For each speech signal, features are extracted from all 25 hidden layers of the SSL models, consisting of the CNN output (first layer) and 24 Transformer layers. Each layer produced a feature matrix comprising 1024-dimensional vectors corresponding to input speech frames, preserving the frame-wise granularity. These SSL-extracted features were integrated into the Kaldi pipeline, replacing the traditional MFCC features. To stabilize the feature distribution before training the DNN model and constructing the KWS system, we applied CMVN.

\section{Results and Discussion}
\label{sec:sec4}
The baseline results using MFCC are presented in Table \ref{tab:performance_metrics}. 
The KWS performance is assessed using metrics proposed by NIST for spoken term detection, including Actual Term Weighted Value (ATWV), Maximum Term Weighted Value (MTWV), probability of false alarm ($P_{\text{fa}}$) and probability of miss ($P_{\text{miss}}$) \cite{fiscus2007results}. This approach aligns with prior studies \cite{pattanayak2022significance, shahnawazuddin2019improving}. For a good KWS performance, TWV values should be closer to 1 \cite{wegmann2013tao} while lower $P_{\text{fa}}$ and $P_{\text{miss}}$ are preferred. For the 10-keyword set, the baseline KWS system achieved ATWV and MTWV scores of -4.827 and -4.674, respectively, with corresponding $P_{\text{fa}}$ and $P_{\text{miss}}$ values of 0.153 and 0.270. For the 20-keyword set, the system produced ATWV and MTWV values of -2.895 and -2.714, alongside $P_{\text{fa}}$ and $P_{\text{miss}}$ rates of 0.181 and 0.267. Similarly, for the 30-keyword set, the baseline system recorded ATWV and MTWV scores of -1.907 and -1.764, with $P_{\text{fa}}$ and $P_{\text{miss}}$ values of 0.176 and 0.256, respectively.
 
\subsection{SSL Layer-wise Features Performance}
\label{ssl-features}
The KWS performance using layer-wise SSL features are illustrated in
figures~\ref{fig:atwv_mtwv} and \ref{fig:pfa_pmiss} for keyword sets of 10, 20 and 30 words. These plots present the ATWV, MTWV, ($P_{\text{fa}}$) and miss ($P_{\text{miss}}$) across all 25 layers of Wav2Vec2, HuBERT and Data2Vec. The best-performing layers for each model and keyword set are summarized in Table~\ref{tab:performance_metrics}.

Across all models, the early layers, layer 0 (the CNN output) and transformer layers 1 to 4 consistently yield highly negative ATWV and MTWV scores. This indicates that low-level spectral representations, while capturing acoustic cues such as formant structures and energy variations, are inadequate for reliable keyword detection in the challenging zero-shot, child-speech setting. For example, HuBERT’s layer 0 yields an ATWV of -1.9072 and MTWV of -1.7251 for the 10 keyword set.

Performance gradually improves across the middle layers, where transformer encoders begin modeling phonetic and syllabic context. HuBERT, shows a clearer gain by layer 10 (ATWV 0.1020, MTWV 0.1406 for 10 keywords). Wav2Vec2 shows a similar trend, with an inflection around layer 9 (ATWV 0.0079, MTWV 0.004), while Data2Vec sees a comparable rise by layer 11 (ATWV -0.0860, MTWV -0.0327).

The highest performance is consistently observed in the later layers (13-22), where the models encode increasingly abstract, semantically rich representations. HuBERT achieves its peak at layer 21 (ATWV 0.0577, MTWV 0.0841 on 10 keywords), followed by a slight decline in later layers. Wav2Vec2 reaches its best at layer 22 (ATWV 0.6917, MTWV 0.7003), while Data2Vec also peaks at layer 22 (ATWV 0.4711, MTWV 0.4711). These top layers, shaped by their respective learning paradigms, contrastive loss in Wav2Vec2, masked prediction in HuBERT and the teacher-student objective in Data2Vec, capture contextualized subword and word-level semantics that are critical for accurate detection. They offer strong true positive rates and minimal false alarms, even under zero-shot conditions. In essence, while low and mid-level representations provide foundational cues, it is the high-level semantic abstractions in the later transformer layers that are pivotal for robust and accurate keyword spotting in children’s speech. 

These results show significant improvement of performance by our proposed system over MFCC based baseline. Statistical analyses including paired t-tests and Wilcoxon signed-rank tests were also performed, confirming that the observed improvements are statistically significant. Thus, while MFCCs remain attractive for their efficiency and simplicity, SSL features present a compelling trade-off: higher computational cost in exchange for significantly improved robustness.

\renewcommand{\arraystretch}{0.8}
\begin{table}[!ht]
    \centering
    \caption{Baseline (MFCCs) KWS performance along with best performing features of SSL layer for Wav2Vec2, HuBERT and Data2Vec models for PFSTAR test set.}
\vspace{0.2cm}
    \resizebox{8cm}{!}{ 
    \begin{tabular}{cccccc}
    \toprule
    \textbf{} & \textbf{} & \textbf{10 keywords} & \textbf{} & \textbf{} & \textbf{} \\
         \toprule
    \textbf{Model} & \textbf{Layer} & \textbf{ATWV} & \textbf{MTWV} & \textbf{$P_{\text{fa}}$} & \textbf{$P_{\text{miss}}$} \\
    \toprule
         Baseline & - & -4.827 & -4.674 & 0.153 & 0.270 \\
         HuBERT & 21 & 0.057 & 0.084 & 0.021 & 0.077 \\
         Wav2Vec2 & 22 & \textbf{0.535} & \textbf{0.535} & \textbf{0.010} & \textbf{0.059 }\\
         Data2Vec & 21 & 0.076 & 0.076 & 0.021 & 0.062 \\
         \bottomrule
         
         \textbf{} & \textbf{} & \textbf{20 keywords} & \textbf{} & \textbf{} & \textbf{} \\
         \toprule
         \textbf{Model} & \textbf{Layer} & \textbf{ATWV} & \textbf{MTWV} & \textbf{$P_{\text{fa}}$} & \textbf{$P_{\text{miss}}$} \\
    \toprule
         Baseline & - & -2.895 & -2.714 &0.181  & 0.267 \\
         HuBERT & 21 & 0.380 & 0.419 & 0.026 & 0.075 \\
         Wav2Vec2 & 22 & \textbf{0.661} & \textbf{0.674} & \textbf{0.013} & \textbf{0.057 }\\
         Data2Vec & 22 & 0.381 & 0.381 & 0.025 & 0.073 \\
         \bottomrule
         
         \textbf{} & \textbf{} & \textbf{30 keywords} & \textbf{} & \textbf{} & \textbf{} \\
         \toprule
         \textbf{Model} & \textbf{Layer} & \textbf{ATWV} & \textbf{MTWV} & \textbf{$P_{\text{fa}}$} & \textbf{$P_{\text{miss}}$} \\
    \toprule
         Baseline & - & -1.907 & -1.764 & 0.176 &0.256  \\
         HuBERT & 21 & 0.457 & 0.483 & 0.030 & 0.071 \\
         Wav2Vec2& 22 & \textbf{0.691} & \textbf{0.700} & \textbf{0.016} & \textbf{0.054} \\
         Data2Vec & 22 & 0.471 & 0.471 & 0.029 & 0.069 \\
         \bottomrule
        
    \end{tabular}
    }
    \vspace{-0.1cm}
    \label{tab:performance_metrics}
\end{table}

\begin{table}[ht]
\renewcommand{\arraystretch}{0.9}
\small
\centering
\caption{KWS performance for the modified PFSTAR test set and the WSJCAM0 augmented training set.}
\label{tab:combined_modification_results}
\resizebox{0.9\linewidth}{!}{%
\begin{tabular}{cccccc}
\toprule
\textbf{Keywords} & \textbf{Modification} 
  & \multicolumn{2}{c}{\textbf{Modified PFSTAR}} 
  & \multicolumn{2}{c}{\textbf{Augmented WSJCAM0}} \\
\cmidrule(lr){3-4} \cmidrule(lr){5-6}
 & 
  & \textbf{ATWV} & \textbf{MTWV} 
  & \textbf{ATWV} & \textbf{MTWV} \\
\midrule
10 &                      & -4.8272 & -4.6743 & -4.8272 & -4.6743 \\
20 & Baseline (no mod)    & -2.8957 & -2.7141 & -2.8957 & -2.7141 \\
30 &                      & -1.9078 & -1.7641 & -1.9078 & -1.7641 \\
\midrule
10 &                      & -5.4383 & -5.2275 & -0.9698 & -0.9698 \\
20 & FM                   & -3.0817 & -2.8592 & -0.9306 & -0.8385 \\
30 &                      & -1.9653 & -1.7908 & -0.4522 & -0.3842 \\
\midrule
10 &                      & -5.1980 & -4.9310 & -0.4388 & -0.4388 \\
20 & PM                   & -2.5218 & -2.3493 & -0.4156 & -0.4236 \\
30 &                      & -1.5617 & -1.4377 & -0.4050 & -0.4050 \\
\midrule
10 &                      & -3.9678 & -3.8403 & -0.3105 & -0.3256 \\
20 & SR                   & -2.4177 & -2.3110 & -0.2934 & -0.2934 \\
30 &                      & -1.5242 & -1.4322 & -0.2362 & -0.2360 \\
\bottomrule
\end{tabular}%
}
\end{table}

\vspace{-0.3cm}

\subsection{Effect of Data Augmentation and Modified Test Set}

To mitigate the acoustic mismatch between adult and children speech in our zero-shot setup, we conducted additional experiments. Three types of speech modifications were applied to the PFSTAR test set, along with augmentation of the WSJCAM0 training set. For pitch and speaking‐rate modifications, we used the RTISI-LA algorithm \cite{RTISILA_2007,kathania_2018_cssp,Kathania2020_interspeech}. Formant modification \cite{kathania2020study} was achieved by systematically shifting the resonant frequencies in the linear-prediction (LP) spectrum. According to prior studies as cited above we applied a pitch modification factor of 0.8, speaking rate factor of 1.2 and a formant modification factor of 0.1 to the PFSTAR test set to mimic the children speech to that of an adult. 

Furthermore, same techniques were applied to augment the WSJCAM0 training set. A pitch modification factor of 1.2, a speaking-rate factor of 0.8, and a formant modification factor of -0.05 were used to transform adult speech to more closely resemble the acoustic characteristics of children’s speech. We then combined these augmented samples with the original WSJCAM0 data. Utilizing the best case scenario of the SSL layer features which is the layer 22 of the Wav2Vec2 model, we trained a KWS system based on these modified test set and augmented training set.

Table \ref{tab:combined_modification_results} presents the KWS performance after these modifications. It can be seen that, modifying the test set by FM and PM provided no significant improvement over the MFCC based baseline. However, applying SR modification did provide improvement over the baseline MFCC based KWS yet still underperformed than our proposed framework. Applying augmentation also showed improved performance over the baseline but still fell short of the performance achieved by our proposed framework.

\subsection{Generalization of the proposed KWS Framework with CMU Kids Corpus}
To further strengthen our study, we extended our experiments to an additional dataset, the CMU Kids Corpus \cite{eskenazi1997cmu}. CMU is a read-speech collection developed at Carnegie Mellon University for the LISTEN tutoring system. This corpus contains speech from 76 American children (24 male, 52 female) aged 6-11 years, comprising 5,180 utterances and approximately 9 hours of speech. For evaluation, we used the best performing features identified in Table \ref{tab:performance_metrics}, layer 22 of the Wav2Vec2 model. However to evaluate the CMU kids speech in a zero-shot scenario, WSJ is not suitable due to the accent mismatch as CMU Kids contains American English and WSJCAM0 contains British English. So we trained our model with MiniLibriSpeech \cite{7178964} dataset, which contains American adult speech. Table \ref{tab:keywords_cmu} shows the keywords used from the CMU dataset for the study.

\renewcommand{\arraystretch}{0.8}
\begin{table}[!htbp]
    \centering
    \footnotesize
    \caption{List of keywords from the CMU dataset used in the study.}
    \label{tab:keywords_cmu}
    \vspace{0.2cm}
   \begin{tabular}{p{2cm} p{5.5cm}} \hline 
         \textbf{No. of Keywords} & \textbf{List of Keywords}\\ \hline
         10 & PEOPLE, LIVE, WATER, USE, CAN, NOT, RAIN, FOOD, EAT, NOW\\ 
        
         20 & PEOPLE, LIVE, WATER, USE, CAN, NOT, RAIN, FOOD, EAT, NOW,  {\color{blue} SAY, NEW, LONG, DUST, AWAY, GROUND, BIG, OUT, GROW, PAPER} \\ 
    
         30 & PEOPLE, LIVE, WATER, USE, CAN, NOT, RAIN, FOOD, EAT, NOW, {\color{blue} SAY, NEW, LONG, DUST, AWAY, GROUND, BIG, OUT, GROW, PAPER,}  {\color{magenta} GOT, MAKE, SOMETIMES, LIGHTNING, ANIMALS, SCIENTISTS, AROUND, WINDS, PLANTS, KEEP} \\ \hline
    \end{tabular}
    
\end{table}

The results, shown in Table~\ref{tab:layerwise_results}, compare the MFCC baseline with the SSL layer features based KWS. The ATWV scores for SSL layer features based KWS are 0.2446, 0.2554 and 0.2986 for 10, 20 and 30 keywords respectively which shows significant improvement over the MFCC based baseline.

\vspace{-0.3cm}
\begin{table}[ht]
\renewcommand{\arraystretch}{0.85}
\small
\centering
\caption{ ATWV and MTWV scores for CMU Kids dataset for MFCC based baseline and SSL features based KWS.}
\vspace{0.3cm}

\begin{tabular}{cccc}
\hline
\textbf{Keywords} & \textbf{Layer} & \textbf{ATWV} & \textbf{MTWV} \\
\hline
10 &  & -0.7389 & -0.6379 \\
20 &  Baseline        & -0.5137 & -0.4270 \\
30 &          & -0.4128 & -0.3385 \\
\hline
10 &   &  0.2446 &  0.2623 \\
20 &   Layer22        &  0.2554 &  0.2863 \\
30 &          &  0.2986 &  0.3057 \\
\hline
\end{tabular}
\label{tab:layerwise_results}
\end{table}

\renewcommand{\arraystretch}{0.9}
\begin{table}[!ht]
     \centering
     \caption{Age group wise analysis of the best performing layer 22 of Wav2vec2 model for PFSTAR test set for 30 keywords set.}
    \vspace{0.2cm}
     \resizebox{6.5cm}{!}{ 
     \begin{tabular}{cccccc}
     \toprule
     \textbf{} &  & \textbf{Age 4-6} &  \textbf{}&  \textbf{}\\
     \toprule
          Model &  ATWV & MTWV&  $P_{\text{fa}}$&  $P_{\text{miss}}$\\
          \hline
          \vspace{2pt}
          Baseline & -6.319&-5.965 &0.573 &0.681 \\
          \vspace{2pt}
          Wav2Vec2 & \textbf{-0.136}&\textbf{-0.020 }&\textbf{0.058}&\textbf{0.083} \\         
          \bottomrule
           \textbf{} &  &\textbf{Age 7-9}&  \textbf{}&  \textbf{}\\
           \toprule
           Model &  ATWV & MTWV&  $P_{\text{fa}}$&  $P_{\text{miss}}$\\
          \hline
          \vspace{2pt}
          Baseline &-1.329 &-1.200 & 0.140 &0.239 \\
          \vspace{2pt}
          Wav2Vec2 & \textbf{0.749}& \textbf{0.749}&\textbf{0.012}&\textbf{0.050} \\
          \bottomrule
          \textbf{} &  &\textbf{Age 10-13} &  \textbf{}&  \textbf{}\\
           \toprule
          Model &  ATWV & MTWV&  $P_{\text{fa}}$&  $P_{\text{miss}}$\\
          \hline
          \vspace{2pt}
          Baseline &-0.827 &-0.827 & 0.099 & 0.118 \\
          \vspace{2pt}
          Wav2Vec2 &\textbf{0.879} &\textbf{0.879} &\textbf{0.006}&\textbf{0.048 }\\
          \bottomrule
     \end{tabular}
     }
     \label{tab:age_class}
 \end{table}

\vspace{-0.4cm}
\subsection{Age wise analysis}
Further, employing the best performing layer, layer 22 from the Wav2Vec2 model, we evaluated KWS performance across three age groups: 4-6 years, 7-9 years, and 10-13 years. We evaluated age-wise analysis using 30 keywords for better gereralisation. The results for 30 keywords set, presented in Table \ref{tab:age_class}, highlight significant differences in performance due to variations in speech characteristics among children for the PFSTAR dataset. The age group 10-13 years achieved the highest scores, with a ATWV of 0.879, MTWV of 0.879, and $P_{\text{fa}}$ and $P_{\text{miss}}$ values of 0.006 and 0.048, respectively. In contrast, the 4-6 age group showed the poorest performance, with an ATWV of -0.136 and MTWV of -0.020 and $P_{\text{fa}}$ and $P_{\text{miss}}$ values of 0.058 and 0.083, reflecting the challenges posed by younger children's speech. The 7-9 age group demonstrated intermediate results, with an ATWV of 0.749, MTWV of 0.749 and $P_{\text{fa}}$ and $P_{\text{miss}}$ values of 0.012 and 0.050, respectively.

Table \ref{tab:age_class_cmu} shows the age wise evaluation on the CMU test set. It can be seen that even for CMU dataset age wise analysis it follows the same trend as for PFSTAR dataset. Lower age group of 6-8 years show poor performance while the higher age group of 9-11 years show improved performance. However for both the age groups our proposed framework significantly improves the KWS performance over the baseline. The poor performance of the lower age groups mirrors prior findings that early childhood speech exhibits high acoustic variability, stemming from incomplete articulatory development, pronunciation inconsistencies, limited phonemic control, and wider pitch, timing, and prosodic fluctuations. Even under these challenging conditions, our zero-shot SSL-based system substantially outperformed the MFCC baseline, underscoring the robustness of SSL representations in adapting to developmental variability. Unlike MFCCs, which capture only surface level spectral features, SSL embeddings encode deeper phonetic and temporal structures that are less affected by age-related variability in pitch, articulation and speaking rate.

\vspace{-0.2cm}
\begin{table}[!ht]
\renewcommand{\arraystretch}{0.85}
     \centering
     \caption{Age group wise analysis of the best performing layer 22 of Wav2vec2 model for CMU test set for 30 keywords set.}
    \vspace{0.2cm}
     \resizebox{6.4cm}{!}{ 
     \begin{tabular}{ccccc}
     \toprule
      \textbf{} &  & \textbf{Age 6-8} &  \textbf{}&  \textbf{}\\
     \toprule
          Model &  ATWV & MTWV&  $P_{\text{fa}}$ &  $P_{\text{miss}}$\\
          \hline
          \vspace{2pt}
          Baseline & -1.126 &-1.096 & 0.213 & 0.329\\
          \vspace{2pt}
          Wav2Vec2 & \textbf{-0.134}&\textbf{-0.127 } & \textbf{0.051}&\textbf{0.068}\\         
          \bottomrule
           
          \textbf{} &  & \textbf{Age 9-11} &  \textbf{}&  \textbf{}\\
     \toprule
          Model &  ATWV & MTWV&  $P_{\text{fa}}$&  $P_{\text{miss}}$\\
          \hline
          \vspace{2pt}
          Baseline &0.154 &0.211 & 0.029&0.038 \\
          \vspace{2pt}
          Wav2Vec2 &\textbf{0.558} &\textbf{0.563} & \textbf{0.004}& \textbf{0.014}\\
          \bottomrule
     \end{tabular}
     }
     \label{tab:age_class_cmu}
 \end{table}
 
\vspace{-0.2cm}
 \subsection{Noise Robustness Analysis of KWS}
To assess the robustness of the proposed system under adverse acoustic conditions, the PFSTAR test set was corrupted with additive noise at varying signal-to-noise ratios (SNRs) ranging from 5 dB to 15 dB, in 5 dB increments. Eight distinct noise types were used, sourced from the NOISEX-92 \cite{varga1993assessment} and MUSAN \cite{snyder2015musan} datasets, to simulate a wide range of real-world scenarios. The evaluation was carried out using both a baseline MFCC-based KWS model and the best-performing layer (layer 22) of the Wav2Vec2 SSL model. The NOISEX-92 dataset, commonly used for benchmarking noise robustness in speech systems, contributed four noise types: babble, factory, white, and Volvo noise. The MUSAN dataset, known for its diversity of samples designed for evaluating speech systems under noisy conditions, provided four additional noise types: thunderstorm (representing low-frequency environmental disturbances), ambulance siren (capturing high-intensity urban sounds), crowd chatter (emulating busy public spaces), and bird chirping (mimicking dynamic natural outdoor environments). This comprehensive noise setup enabled a rigorous evaluation of the KWS system’s performance across varied acoustic challenges.

\renewcommand{\arraystretch}{0.85}
\begin{table}[!ht]
    \centering
    \small
    \caption{ATWV scores for 10, 20 and 30 keywords sets under different noise conditions and SNR levels for MFCC (baseline) and Wav2Vec2 layer 22 features for PFSTAR test set.}
    \label{tab:noise kws_pfstar}
    \vspace{0.2cm}
    \resizebox{8.5cm}{!}{
    \begin{tabular}{@{}lcccccccccc@{}}
        \toprule
        \textbf{Noise Type} & \textbf{SNR (dB)} & \multicolumn{2}{c}{\textbf{10 Keywords}} & \multicolumn{2}{c}{\textbf{20 Keywords}} & \multicolumn{2}{c}{\textbf{30 Keywords}} \\
        \cmidrule(lr){3-4} \cmidrule(lr){5-6} \cmidrule(lr){7-8}
        & & \textbf{MFCC} & \textbf{Layer 22} & \textbf{MFCC} & \textbf{Layer 22} & \textbf{MFCC} & \textbf{Layer 22} \\
        \midrule
        \multirow{3}{*}{Babble} 
        & 5  & -9.184 & -2.111 & -8.304 & -0.677 & -8.012 & -0.193 \\
        & 10 & -6.285 & -0.676 & -6.176 &  0.070 & -6.083 &  0.321 \\
        & 15 & -5.519 & -0.428 & -5.236 &  0.170 & -5.117 &  0.352 \\
        \midrule
        \multirow{3}{*}{Factory} 
        & 5  & -12.366 & -3.435 & -12.254 & -1.525 & -12.118 & -0.912 \\
        & 10 & -11.550 & -0.886 & -11.462 & -0.108 & -11.015 &  0.116 \\
        & 15 & -10.113 & -0.374 & -10.106 &  0.187 & -10.093 &  0.366 \\
        \midrule
        \multirow{3}{*}{Volvo} 
        & 5  & -4.845 &  0.039 & -4.716 &  0.408 & -4.711 &  0.519 \\
        & 10 & -4.506 &  0.204 & -4.494 &  0.491 & -4.412 &  0.557 \\
        & 15 & -4.222 &  0.204 & -4.164 &  0.491 & -4.114 &  0.567 \\
        \midrule
        \multirow{3}{*}{White} 
        & 5  & -17.137 & -2.903 & -16.990 & -1.270 & -16.885 & -0.835 \\
        & 10 & -13.203 & -0.959 & -13.198 & -0.202 & -13.101 &  0.041 \\
        & 15 & -10.352 & -0.737 & -10.137 &  0.021 & -10.101 &  0.261 \\
        \midrule
        \multirow{3}{*}{Ambulance Siren} 
        & 5  & -7.252 & -3.834 & -7.210 & -1.743 & -7.127 & -1.229 \\
        & 10 & -6.321 & -2.392 & -6.301 & -0.813 & -6.286 & -0.312 \\
        & 15 & -4.896 & -0.181 & -4.884 &  0.303 & -4.811 &  0.433 \\
        \midrule
        \multirow{3}{*}{Crowd} 
        & 5  & -9.064 & -4.300 & -9.007 & -1.924 & -9.001 & -1.379 \\
        & 10 & -8.898 & -2.198 & -8.630 & -0.705 & -8.611 & -0.281 \\
        & 15 & -6.125 & -1.200 & -6.111 & -0.202 & -6.102 &  0.088 \\
        \midrule
        \multirow{3}{*}{Thunder Storm} 
        & 5  & -9.247 & -0.157 & -9.118 &  0.334 & -9.029 &  0.421 \\
        & 10 & -7.739 & -0.264 & -7.564 &  0.284 & -7.413 &  0.430 \\
        & 15 & -6.540 & -0.043 & -6.302 &  0.368 & -6.115 &  0.513 \\
        \midrule
        \multirow{3}{*}{Birds Chirping} 
        & 5  & -11.196 & -3.748 & -11.025 & -1.693 & -11.011 & -1.053 \\
        & 10 & -9.961 & -1.122 & -9.866 & -0.199 & -9.305 &  0.028 \\
        & 15 & -8.159 & -1.813 & -8.118 & -0.547 & -8.027 & -0.139 \\
        \bottomrule
    \end{tabular}}
\end{table}

The impact of noise on the PFSTAR test set is presented in Table~\ref{tab:noise kws_pfstar}. The results indicate that, even under noisy conditions, the SSL-based KWS system significantly outperforms the MFCC-based baseline. Similarly, Table~\ref{tab:noise_kws_cmu} displays the performance on the CMU dataset when subjected to comparable noise conditions. The CMU results follow a similar trend, with the SSL-based approach consistently providing substantial improvements over the baseline.

These findings suggest that features derived from SSL models are inherently more robust to noise, owing to their deep Transformer architectures that capture long-range dependencies and contextual information. Additionally, pretraining on large, acoustically diverse corpora enables these models to learn to disregard irrelevant or transient acoustic variations.

\vspace{-0.2cm}
\begin{table}[!ht]
    \renewcommand{\arraystretch}{0.85}

    \centering
    \small
    \caption{ATWV scores for 10, 20 and 30 keywords sets under different noise conditions and SNR levels for MFCC (baseline) and Wav2Vec2 layer 22 features for CMU test set.}
    \label{tab:noise_kws_cmu}
    \vspace{0.2cm}
    \resizebox{8.5cm}{!}{
    \begin{tabular}{@{}lcccccccccc@{}}
        \toprule
        \textbf{Noise Type} & \textbf{SNR (dB)} & \multicolumn{2}{c}{\textbf{10 Keywords}} & \multicolumn{2}{c}{\textbf{20 Keywords}} & \multicolumn{2}{c}{\textbf{30 Keywords}} \\
        \cmidrule(lr){3-4} \cmidrule(lr){5-6} \cmidrule(lr){7-8}
        & & \textbf{MFCC} & \textbf{Layer 22} & \textbf{MFCC} & \textbf{Layer 22} & \textbf{MFCC} & \textbf{Layer 22} \\
        \midrule
        \multirow{3}{*}{Babble} 
        & 5  & -12.165 & -3.963 & -11.636&-2.306 &-12.060 & -1.968\\
        & 10 & -10.639 & -2.147 &-10.115 & -1.658 &-9.820 &-1.493 \\
        & 15 & -8.336 &-1.006 &-8.306 & -1.151 &-8.124 & -1.003 \\
        \midrule
        \multirow{3}{*}{Factory} 
        & 5  & -14.908 & -4.968 & -13.230 & -4.192 & -12.604 & -3.560\\
        & 10 & -12.709 & -3.610 & -12.200 & -3.190 & -11.564 &-3.190 \\
        & 15 & -10.850  & -3.110& -10.180 & -3.006 & -10.230& -2.907 \\
        \midrule
        \multirow{3}{*}{Volvo} 
        & 5  & -6.110 & -0.963 & -6.071 & -0.128 &-6.040 &0.198 \\
        & 10 & -5.908 &-0.682 & -5.859 &-0.079 & -5.762 &0.278 \\
        & 15 & -5.699 &-0.314 & -5.110 & 0.150&-5.690 & 0.319\\
        \midrule
        \multirow{3}{*}{White} 
        & 5  & -19.090 & -4.989 &-18.806 & -4.890 & -18.101 &-4.827 \\
        & 10 & -18.349  & -4.780 & -18.107 & -4.207 & -18.102 & -4.305 \\
        & 15 & -15.560 & -4.206 &-15.401 & -4.178&-15.119 & -4.193\\
        \midrule
        \multirow{3}{*}{Ambulance Siren} 
        & 5  & -9.768 & -3.631 & -9.567 &-3.504 & -9.198 &-3.110 \\
        & 10 & -9.137 &-3.161 &-9.365 &-3.107 & -9.117& -3.012\\
        & 15 & -8.961 &-2.947 & -9.106&-2.881 &-8.923 & -2.813\\
        \midrule
        \multirow{3}{*}{Crowd} 
        & 5  & -10.369 &5.861 & -10.305 &-4.106 & -10.203 & -4.096 \\
        & 10 & -8.364 &-4.114 & -8.111 &-4.002 & -8.021& -3.993 \\
        & 15 &-7.664  & -3.209&-7.403 &-3.193 & -7.019&-3.100 \\
        \midrule
        \multirow{3}{*}{Thunder Storm} 
        & 5  & -10.667 &-1.963 & -10.113&-1.855 & -10.103 &-1.772 \\
        & 10 & -8.139 &-1.668 &-8.256 & -1.302&-8.003 & -1.521\\
        & 15 & -7.628  &-0.610 & -7.164& -0.561 &-7.310 & -0.314\\
        \midrule
        \multirow{3}{*}{Birds Chirping} 
        & 5  & 12.725 &-4.213 &-12.460 &-4.016 & -12.061 & -4.003\\
        & 10 &-10.240 &-3.690 & -10.113& 3.551&-10.098 &-3.463\\
        & 15 & -9.560 &-2.450 & -9.101& -2.439& -8.973 &-2.304 \\
        \bottomrule
    \end{tabular}}
\end{table}

\vspace{-0.3cm}
\subsection{Comparison with previous studies}
This section presents a comparison of the proposed KWS framework with three prior studies that follow a similar experimental setup, specifically those using the PFSTAR dataset as the test set. Table~\ref{tab:compared_work} summarizes the results of these studies alongside the proposed approach. Shahnawazuddin et al. \cite{shahnawazuddin2019improving} proposed prosody modification-based data augmentation techniques to enhance KWS performance in children’s speech. In \cite{pattanayak2021pitch}, a pitch-robust acoustic feature, Mel-spaced Single Frequency Average Log Envelope (MSSF-ALE), was introduced. Building on this, Pattanayak et al. \cite{pattanayak2022significance} proposed a pitch-robust feature extraction technique using Single Frequency Filtering (SFF) to derive Mel-spaced amplitude envelopes, termed MS-SFF-CC. Compared to these approaches, the proposed KWS system, which utilizes features extracted from layer 22 of the Wav2Vec2 model, achieves superior performance on the same test dataset, highlighting the effectiveness of self-supervised learning representations for keyword spotting in children’s speech.

\vspace{-0.3cm}
\renewcommand{\arraystretch}{0.9}
\begin{table}[!ht]
    \centering
    \small
    \caption{ Comparative performance of the proposed framework with previous studies for 10 and 20 keywords set.}
    \vspace{0.2cm}
    \label{tab:compared_work}
    \resizebox{8.0cm}{!}{ 
    \begin{tabular}
    {p{2.2cm} p{4cm}  p{0.9cm} p{0.9cm}} 
    \toprule
       \textbf{Author} & \textbf{Methodology }  &  \textbf{ATWV (10 kw)} &  \textbf{ATWV (20 kw)} \\ \toprule
        
        Pattanayak et al. (2022) \cite{pattanayak2022significance} & A pitch independent feature extraction technique employing SFF is proposed.  & 0.422  & 0.353\\ \hline

         Pattanayak et al. (2021) \cite{pattanayak2021pitch} & A pitch robust acoustic feature based on single frequency filtering (SFF) is proposed.  &  0.229 & - \\
         \hline
          Shahnawazuddin et al. (2019) \cite{shahnawazuddin2019improving} & Prosody modification based data augmentation  were explored. & 0.309 & 0.174 \\ \hline
        Proposed & SSL features using Wav2Vec2 model for layer 22  & \textbf{0.535}  &\textbf{0.661} \\ \bottomrule

    \end{tabular}}
\end{table}

\vspace{-0.5cm}
\section{Conclusion and Future Scope}
\label{sec:sec5}
\vspace{-0.2cm}
In this paper, we introduced a Zero-Shot KWS for children's speech using the layer-wise features from three state-of-the-art pre-trained SSL models: Wav2Vec2, HuBERT and Data2Vec. The KWS system was trained on WSJCAM0 adult speech data and evaluated on PFSTAR children’s speech, highlighting its zero-shot capability. Our results demonstrate that KWS performance improves with an increasing number of keywords. Layer 22 of Wav2Vec2 model performed the best with  ATWV of 0.691, MTWV of 0.700 and $P_{\text{fa}}$ and $P_{\text{miss}}$ values of 0.016 and 0.054, respectively for the 30 keywords set. Further analysis across different age groups revealed that KWS performance improves with increasing age, highlighting challenges, particularly with younger children. We carried out additional experiments for noisy data for the best performing Wav2vec2 layer by adding noise from NOISEX-92 and MUSAN dataset into the test set. The results demonstrated significant improvement over traditional MFCC based KWS system thereby emphasizing the system's noise robustness. A comparative evaluation with prior studies shows that the proposed approach outperforms existing methods on the same test dataset. To validate the generalizability of our framework, we conducted additional zero-shot experiments using the CMU Kids dataset, observing consistent trends in performance. These findings support the hypothesis that SSL-derived features capture robust and invariant representations, making them highly effective for keyword spotting tasks in acoustically challenging and resource-constrained scenarios. Overall, the study demonstrates the potential of SSL-based models for enhanced zero-shot KWS in children’s speech.

Future extensions of this framework include the integration of SSL embeddings with visual lip-movement features, aiming to enhance keyword spotting performance under severe noise conditions.

\vspace{-0.4cm}

\bibliographystyle{elsarticle-num}
\bibliography{main}

\begin{thebibliography}{10}
\expandafter\ifx\csname url\endcsname\relax
  \def\url#1{\texttt{#1}}\fi
\expandafter\ifx\csname urlprefix\endcsname\relax\def\urlprefix{URL }\fi
\expandafter\ifx\csname href\endcsname\relax
  \def\href#1#2{#2} \def\path#1{#1}\fi

\bibitem{weintraub1995lvcsr}
M.~Weintraub, Lvcsr log-likelihood ratio scoring for keyword spotting, in: International Conference on Acoustics, Speech, and Signal Processing, Vol.~1, IEEE, 1995, pp. 297--300.

\bibitem{szoke2005comparison}
I.~Sz{\"o}ke, P.~Schwarz, P.~Matejka, L.~Burget, M.~Karafi{\'a}t, M.~Fapso, J.~Cernock{\`y}, Comparison of keyword spotting approaches for informal continuous speech., in: Interspeech, 2005, pp. 633--636.

\bibitem{rose1990hidden}
R.~C. Rose, D.~B. Paul, A hidden markov model based keyword recognition system, in: International conference on acoustics, speech, and signal processing, IEEE, 1990, pp. 129--132.

\bibitem{can2011lattice}
D.~Can, M.~Saraclar, Lattice indexing for spoken term detection, IEEE Transactions on Audio, Speech, and Language Processing 19~(8) (2011) 2338--2347.

\bibitem{hwang2015online}
K.~Hwang, M.~Lee, W.~Sung, \href{https://api.semanticscholar.org/CorpusID:6680389}{Online keyword spotting with a character-level recurrent neural network}, ArXiv abs/1512.08903 (2015).
\newline\urlprefix\url{https://api.semanticscholar.org/CorpusID:6680389}

\bibitem{lopez2021deep}
I.~L{\'o}pez-Espejo, Z.-H. Tan, J.~H. Hansen, J.~Jensen, Deep spoken keyword spotting: An overview, IEEE Access 10 (2021) 4169--4199.

\bibitem{mazumder2021few}
M.~Mazumder, C.~R. Banbury, J.~Meyer, P.~Warden, V.~J. Reddi, \href{https://api.semanticscholar.org/CorpusID:233025251}{Few-shot keyword spotting in any language}, in: Interspeech, 2021.
\newline\urlprefix\url{https://api.semanticscholar.org/CorpusID:233025251}

\bibitem{berg2021keyword}
A.~Berg, M.~O’Connor, M.~T. Cruz, Keyword transformer: A self-attention model for keyword spotting, in: Interspeech 2021, 2021, pp. 4249--4253.
\newblock \href {https://doi.org/10.21437/Interspeech.2021-1286} {\path{doi:10.21437/Interspeech.2021-1286}}.

\bibitem{warden2018speech}
P.~Warden, Speech commands: A dataset for limited-vocabulary speech recognition, arXiv preprint arXiv:1804.03209 (2018).

\bibitem{hussain2022multi}
S.~Hussain, V.~Nguyen, S.~Zhang, E.~Visser, Multi-task voice activated framework using self-supervised learning, in: IEEE International Conference on Acoustics, Speech and Signal Processing (ICASSP), 2022, pp. 6137--6141.

\bibitem{mork2024noise}
J.~M{\o}rk, H.~S. Bovbjerg, G.~Kiss, Z.-H. Tan, Noise-robust keyword spotting through self-supervised pretraining, arXiv preprint arXiv:2403.18560 (2024).

\bibitem{baevski2020wav2vec}
A.~Baevski, Y.~Zhou, A.~Mohamed, M.~Auli, wav2vec 2.0: A framework for self-supervised learning of speech representations, Advances in neural information processing systems 33 (2020) 12449--12460.

\bibitem{chen2022wavlm}
S.~Chen, C.~Wang, Z.~Chen, Y.~Wu, S.~Liu, Z.~Chen, J.~Li, N.~Kanda, T.~Yoshioka, X.~Xiao, et~al., Wavlm: Large-scale self-supervised pre-training for full stack speech processing, IEEE Journal of Selected Topics in Signal Processing 16~(6) (2022) 1505--1518.

\bibitem{zhu2024ge2e}
P.~Zhu, J.~W. Bartel, D.~Agarwal, K.~Partridge, H.~J. Park, Q.~Wang, Ge2e-kws: Generalized end-to-end training and evaluation for zero-shot keyword spotting, in: 2024 IEEE Spoken Language Technology Workshop (SLT), IEEE, 2024, pp. 999--1006.

\bibitem{huang2023building}
R.~Huang, M.~Wiesner, L.~P. Garcia-Perera, D.~Povey, J.~Trmal, S.~Khudanpur, Building keyword search system from end-to-end asr systems, in: ICASSP 2023-2023 IEEE International Conference on Acoustics, Speech and Signal Processing (ICASSP), IEEE, 2023, pp. 1--5.

\bibitem{radford2023robust}
A.~Radford, J.~W. Kim, T.~Xu, G.~Brockman, C.~McLeavey, I.~Sutskever, Robust speech recognition via large-scale weak supervision, in: International conference on machine learning, PMLR, 2023, pp. 28492--28518.

\bibitem{sinha2024effect}
A.~Sinha, M.~Singh, S.~R. Kadiri, M.~Kurimo, H.~K. Kathania, Effect of speech modification on wav2vec2 models for children speech recognition, in: International Conference on Signal Processing and Communications (SPCOM), IEEE, 2024, pp. 1--5.

\bibitem{Li2024AnalysisOS}
J.~Li, M.~A. Hasegawa-Johnson, N.~L. McElwain, \href{https://api.semanticscholar.org/CorpusID:267627009}{Analysis of self-supervised speech models on children’s speech and infant vocalizations}, IEEE International Conference on Acoustics, Speech, and Signal Processing Workshops (ICASSPW) (2024) 550--554.
\newline\urlprefix\url{https://api.semanticscholar.org/CorpusID:267627009}

\bibitem{yu2023few}
M.~Yu, X.~Jin, B.~Wan, G.~Wang, A few-shot speech keyword spotting method based on self-supervise learning, in: 2023 16th International Congress on Image and Signal Processing, BioMedical Engineering and Informatics (CISP-BMEI), IEEE, 2023, pp. 1--5.

\bibitem{jacobs2024multilingual}
C.~Jacobs, Multilingual acoustic word embeddings for zero-resource languages, arXiv preprint arXiv:2401.10543 (2024).

\bibitem{9585401}
W.-N. Hsu, B.~Bolte, Y.-H.~H. Tsai, K.~Lakhotia, R.~Salakhutdinov, A.~Mohamed, Hubert: Self-supervised speech representation learning by masked prediction of hidden units, IEEE/ACM Transactions on Audio, Speech, and Language Processing.

\bibitem{baevski2022data2vec}
A.~Baevski, W.-N. Hsu, Q.~Xu, A.~Babu, J.~Gu, M.~Auli, Data2vec: A general framework for self-supervised learning in speech, vision and language, in: International conference on machine learning, PMLR, 2022, pp. 1298--1312.

\bibitem{povey2011kaldi}
D.~Povey, A.~Ghoshal, G.~Boulianne, L.~Burget, O.~Glembek, N.~Goel, M.~Hannemann, P.~Motlicek, Y.~Qian, P.~Schwarz, et~al., The kaldi speech recognition toolkit, in: IEEE 2011 workshop on automatic speech recognition and understanding, IEEE Signal Processing Society.

\bibitem{robinson1995wsjcamo}
T.~Robinson, J.~Fransen, D.~Pye, J.~Foote, S.~Renals, Wsjcamo: a british english speech corpus for large vocabulary continuous speech recognition, in: 1995 International Conference on Acoustics, Speech, and Signal Processing, Vol.~1, IEEE, pp. 81--84.

\bibitem{batliner2005pf_star}
A.~Batliner, M.~Blomberg, S.~D'Arcy, D.~Elenius, D.~Giuliani, M.~Gerosa, C.~Hacker, M.~Russell, S.~Steidl, M.~Wong, The pf\_star children's speech corpus (2005).

\bibitem{fiscus2007results}
J.~G. Fiscus, J.~Ajot, J.~S. Garofolo, G.~Doddingtion, Results of the 2006 spoken term detection evaluation, in: Proc. sigir, Vol.~7, 2007, pp. 51--57.

\bibitem{pattanayak2022significance}
B.~Pattanayak, G.~Pradhan, Significance of single frequency filter for the development of children's kws system., in: INTERSPEECH, 2022, pp. 3183--3187.

\bibitem{shahnawazuddin2019improving}
S.~Shahnawazuddin, K.~Maity, G.~Pradhan, Improving the performance of keyword spotting system for children's speech through prosody modification, Digital Signal Processing 86 (2019) 11--18.

\bibitem{wegmann2013tao}
S.~Wegmann, A.~Faria, A.~Janin, K.~Riedhammer, N.~Morgan, The tao of atwv: Probing the mysteries of keyword search performance, in: 2013 IEEE Workshop on Automatic Speech Recognition and Understanding, IEEE, pp. 192--197.

\bibitem{RTISILA_2007}
X.~{Zhu}, G.~T. {Beauregard}, L.~L. {Wyse}, Real-time signal estimation from modified short-time fourier transform magnitude spectra, IEEE Transactions on Audio, Speech, and Language Processing (July 2007).

\bibitem{kathania_2018_cssp}
H.~K. {Kathania}, W.~{Ahmad}, S.~{Shahnawazuddin}, A.~B. {Samaddar}, Explicit pitch mapping for improved children’s speech recognition, Circuits, Systems, and Signal Processing (2018).

\bibitem{Kathania2020_interspeech}
H.~Kathania, M.~Singh, T.~Grósz, M.~Kurimo, Data augmentation using prosody and false starts to recognize non-native children's speech, in: Proc. Interspeech 2020, 2020.

\bibitem{kathania2020study}
H.~K. Kathania, S.~R. Kadiri, P.~Alku, M.~Kurimo, Study of formant modification for children asr, in: ICASSP 2020-2020 IEEE International Conference on Acoustics, Speech and Signal Processing (ICASSP), IEEE, 2020, pp. 7429--7433.

\bibitem{eskenazi1997cmu}
M.~Eskenazi, J.~Mostow, D.~Graff, The cmu kids corpus, Linguistic Data Consortium 11 (1997).

\bibitem{7178964}
V.~Panayotov, G.~Chen, D.~Povey, S.~Khudanpur, Librispeech: An asr corpus based on public domain audio books, in: 2015 IEEE International Conference on Acoustics, Speech and Signal Processing (ICASSP), 2015, pp. 5206--5210.
\newblock \href {https://doi.org/10.1109/ICASSP.2015.7178964} {\path{doi:10.1109/ICASSP.2015.7178964}}.

\bibitem{varga1993assessment}
A.~Varga, H.~J. Steeneken, Assessment for automatic speech recognition: Ii. noisex-92: A database and an experiment to study the effect of additive noise on speech recognition systems, Speech communication 12~(3) (1993) 247--251.

\bibitem{snyder2015musan}
D.~Snyder, G.~Chen, D.~Povey, Musan: A music, speech, and noise corpus, arXiv preprint arXiv:1510.08484 (2015).

\bibitem{pattanayak2021pitch}
B.~Pattanayak, G.~Pradhan, Pitch-robust acoustic feature using single frequency filtering for children’s kws, Pattern Recognition Letters 150 (2021) 183--188.

\end{thebibliography}

\end{document}